# Exchange bias in Fe/Ir$_{20}$Mn$_{80}$ bilayers: Role of spin-glass like interface and 'bulk' antiferromagnet spins

Sagarika Nayak, Palash Kumar Manna, Thiruvengadam Vijayabaskaran, Braj Bhusan Singh and Subhankar Bedanta*

Laboratory for Nanomagnetism and Magnetic Materials (LNMM), School of Physical Sciences, National Institute of Science Education and Research (NISER), HBNI, P.O.- Bhimpur Padanpur, Via –Jatni, 752050, India

Email: Subhankar Bedanta - sbedanta@niser.ac.in

* Corresponding author

Abstract

We have performed magnetic measurements like temperature ($T$), cooling field ($H_{FC}$) dependence of exchange bias (EB) and training effect to investigate the magnetic nature of the interface of the Fe/Ir$_{20}$Mn$_{80}$ systems. Thin film bilayer samples of different thicknesses of Ir$_{20}$Mn$_{80}$ have been prepared by dc magnetron sputtering at room temperature. The variation of exchange bias field ($H_{EB}$) with the increase in thickness of Ir$_{20}$Mn$_{80}$ predicts the antiferromagnet (AFM) 'bulk' spins contribution to EB. Exponential decay of $H_{EB}$ and coercive field ($H_C$) with temperature reveals the presence of spin glass (SG) like interface. Also, the decrease of $H_{EB}$ with increasing $H_{FC}$ confirms the SG like frustration at the interface. Further, the fitting of training effect experimental data envisages the presence of frozen and rotatable spins at the magnetically frustrated interface of these EB systems.

## Introduction

The exchange coupling at the interface of ferromagnet (FM)/AFM system [1] develops a unidirectional anisotropy [2-4] in the FM layer, which leads to horizontal shift [5,6] and broadening in the hysteresis loop [7]. Although, EB as an interfacial interaction between FM and AFM is widely studied [8] but the 'bulk' spin contribution to EB is still not fully established [9]. However, EB effects have been widely explored for the fundamental understanding and for its technological applications in constructing spintronic based devices such as spin valves, and magnetic random-access memories (MRAMs) etc [6,9-12].

The presence of interfacial spins and their behaviour play critical role for the origin of EB and the training effect where EB decreases with number of cycles of hysteresis loop measurement. It is reported in literature that the interface of the FM and AFM phase can be spin glass (SG) like disorder [12], compensated AFM order or uncompensated FM order. The structural disorder [12], interface roughness [13], chemical intermixing [13], interdiffusion [13] and lack of structural periodicity [12] might be a reason for this glassy state at the interface. The interface is assumed to play a key role in EB, however the AFM 'bulk' can also sometimes play an important role [10]. The $H_{EB}$ depends primarily on the defects present at the FM/AFM interface but rather on the presence of the defects in the 'bulk' part of the AFM [14]. Further, it is found from a number of experimental evidences that the $H_{EB}$ depends on the AFM layer thickness [15-17] which exists even for higher thicknesses of AFM confirms the contribution of spins from 'bulk' part of the AFM to the EB [18-22]. It is found by the various reports that EB still exists by inserting a non-magnetic layer in between FM and AFM layer [10]. Therefore, the understating the role of the various types of interfacial spins and the 'bulk' AFM spins may lead to unveil the origin of the EB.

In the present study we report the magnetic nature of the interface from various magnetic measurements. In order to probe the role of 'bulk' spins we varied the thickness of $Ir_{20}Mn_{80}$ layer for the fixed Fe layer thickness. The trend of $H_{EB}$ and $H_C$ with temperature and $H_{FC}$ indicates the presence of SG like interface in the FM/AFM system. From the fitting of the training effect experimental data, we found that both the frozen and rotatable interfacial frustrated spins contribute to the EB. The rotatable interfacial spins are found to be relax 8 times faster than the frozen interfacial spins.

Table 1: the sample nomenclature and structure for all the samples.

| Sample name | Sample structure |
| --- | --- |
| S1 | Si (100)/Ta (1 nm)/Cu (10 nm)/IrMn (10 nm)/Fe (10 nm)/Cu (3 nm) |
| S2 | Si (100)/Ta (1 nm)/Cu (10 nm)/IrMn (5 nm)/Fe (10 nm)/Cu (3 nm) |
| S3 | Si (100)/Ta (1 nm)/Cu (10 nm)/IrMn (3 nm)/Fe (10 nm)/Cu (3 nm) |

## Experimental Details

We have deposited Fe/Ir$_{20}$Mn$_{80}$ bilayers on Si (100) substrate having native oxide layer by dc magnetron sputtering at room temperature. Sample nomenclature and structure of all the samples are given in table I. Cu of 10 nm thickness has been deposited on top of Ta of 1 nm thickness as seed layer. To avoid oxidation, Cu of 3 nm thickness has been deposited as capping layer. All the thin film layers have been fabricated in a high vacuum chamber manufactured by Mantis deposition Ltd., UK. The base pressure was better than $3\times10^{-8}$ mbar and the Ar working pressure was $5\times10^{-8}$ mbar. The magnetic measurements such as training effect, $H_{FC}$ and temperature dependence of EB were performed using magnetic property measuring system (MPMS 3) manufactured by Quantum design. Grazing incidence x-ray diffraction (GIXRD) measurement was done with x-ray diffractometer from Rigaku equipped with Cu-K$_\alpha$ x-ray source.

## Results and Discussion

Samples S1-S3 have diffraction peaks correspond to both Fe and IrMn layers at diffraction angle $2\theta$ of 41.9°, 43.2° and 43.5° found in the GIXRD data for all the bilayer thin films (Figure S1 in supplementary material). Fe (111) peak with cubic face-centred crystal structure is found for all the samples. Along with Fe (111) peak, all the samples have IrMn (111) peak with tetragonal primitive crystal structure.

Figure 1 (a) and (b) depict the trend of $H_{EB}$ and $H_C$ vs T of the samples S1-S3. The values of $H_{EB}$ and $H_C$ are taken from the hysteresis loops measured at each temperature after cooling from 400 K in the presence of 1 Tesla (T) magnetic field. It has been found from literatures that in the conventional FM/AFM system the maximum value in HEB occurs at minimum value in HC in a temperature dependent measurement [8]. However, in a system with SG like interface the values of $H_{EB}$ and $H_C$ are found to follow a similar trend [fig. 1(b) and (c)]. We observed exponential decay of $H_{EB}$ and $H_C$ with increasing temperature in all the samples. The exponential decay of the $H_{EB}$ with increasing temperature is due to the increase in long range AFM ordering and hence the increase in the exchange anisotropy [23]. But we observed both the $H_{EB}$ and $H_C$ are decreasing with increasing temperature. This type of decay for both the $H_{EB}$ and $H_C$ has been generally observed in magnetically frustrated system. As in this study the blocking temperature is too high, the interfacial frustrated disorder state might be a reason behind this decay behaviour in the $H_{EB}$, $H_C$ vs $T$ plots [24]. If the interface of the FM/AFM system is not frustrated then we may not observe this type of decay of $H_{EB}$ and $H_C$ with temperature. The blocking temperature at which the EB vanishes is found to be ~ 400 K for all the samples. At temperatures above 50 K the $H_{EB}$ gradually decreases with temperature due to thermal excitations [7]. Also, the interfacial spins become uncorrelated due to thermal energy [12]. Therefore, at higher temperatures the interfacial AFM spins under the polarizing action of the 'bulk' AFM contribute to the $H_{EB}$ and low anisotropy interfacial spins contribute to the $H_C$. We have found a sudden rise of $H_{EB}$ and $H_C$ at lower temperatures below 50 K possibly due to freezing of spin glass like interfacial spins [7]. Our observations are corroborated with findings of the Fulara *et al.* [7] in the Fe/Ir$_{20}$Mn$_{80}$ layer structures. Thus, this temperature dependent study elucidates the presence of SG like interface in these EB systems.

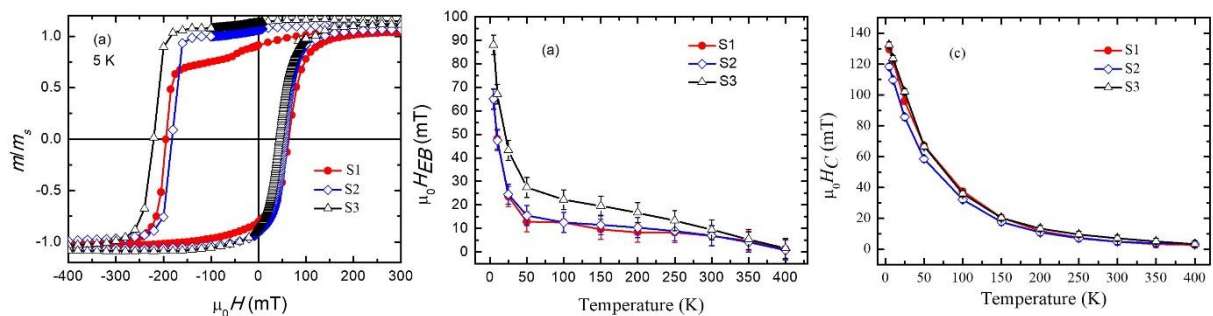

Figure 1: shows the hysteresis loops (a) variation of $H_{EB}$ (a) and $H_C$ (b) vs temperature ($T$) for all the samples.

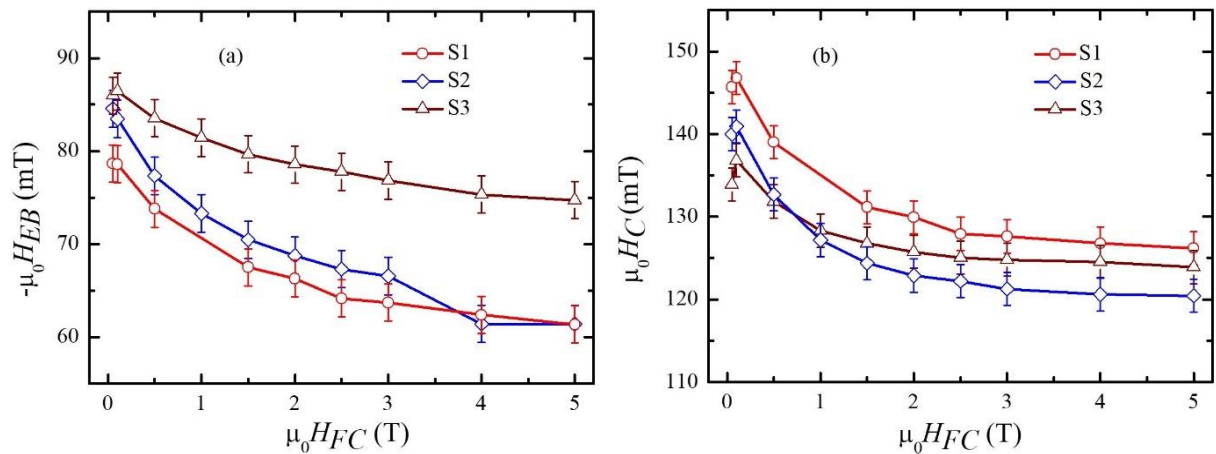

Figure 2: shows the cooling field (HFC) dependence of (a) $H_{EB}$, (b) $H_C$ for samples S1-S3.

Figure 2 (a) and (b) show the variation of $H_{EB}$ and $H_C$ with $H_{FC}$ for samples S1-S3. We have cooled our bilayer samples from 400 K to 5 K in presence of different magnetic fields (500 Oe to 5 T). In conventional FM/AFM system, the $H_{EB}$ increases with increasing $H_{FC}$ [25]. More numbers of spins are getting pinned in the $H_{FC}$ direction and that increases with increase in the $H_{FC}$. Again, the cooling field acts on the 'bulk' of the AFM to induce magnetization in the AFM [14]. However, in the FM/SG system the $H_{EB}$ decreases with increasing HFC [26] due to random interface effect [14]. The cooling field acts on the 'bulk' of the SG to induce some magnetization [14]. But, the random interaction in the 'bulk' and interface of FM/SG systems gives rise to the cancellation of the magnetization

[14]. In our $Fe/Ir_{20}Mn_{80}$ system, we observed monotonous decrease of $H_{EB}$ and $H_C$ with $H_{FC}$. The spin glass like interaction at the interface of Fe and $Ir_{20}Mn_{80}$ might be a reason for this type of behaviour of $H_{EB}$ and $H_C$ with $H_{FC}$.

Figure 3 shows the AFM thickness dependence of EB at different cooling fields. In this study, we found that the $H_{EB}$ is decreasing with increasing the thickness of AFM layer from 3 to 5 nm after this it levels off. In literature, it is found that there is a critical thickness of AFM layer where the onset of EB occurs [8]. It is reported that in sputter deposited thin film samples the critical thickness is above ~ 5 nm [27]. the critical thickness, $H_{EB}$ continues to increase and then a peak in $H_{EB}$ is found [8]. After the peak a decrease in $H_{EB}$ is found followed by saturation [8]. However, in our bilayer samples the AFM thickness is more than the critical thickness. Similar results were found in literature with the explanation of formation of partial domain wall in the AFM parallel to the interface [8]. The reason for this type of AFM thickness dependence may be due to the microstructural changes of the AFM layer as one kind of orientation is not stable above some thickness [28]. The AFM thickness dependence behaviour has also been explained theoretically that there may be the change in AFM domain structure with the thickness of AFM and therefore the variation of $H_{EB}$ with thickness of AFM has been observed [28,29]. Thus, the thickness dependence study gives an overview of the contribution of volume part of the AFM to the EB.

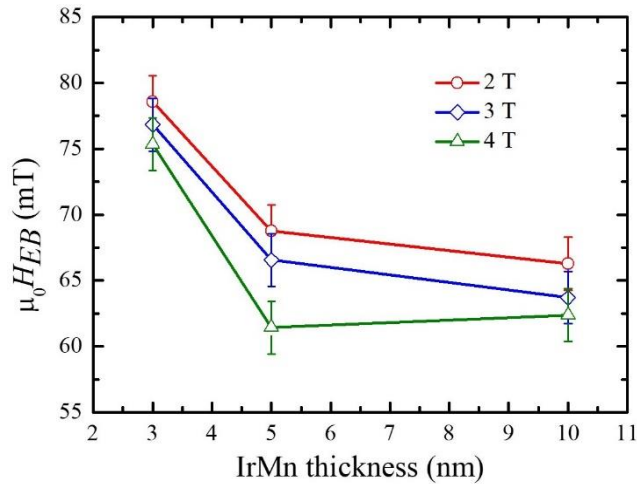

Figure 3: shows the IrMn thickness dependence of exchange bias for cooling field of 2, 3 and 4 T.

Fig 4 (a)-(c) show the M-H loops taken after field cooling the system at 1 T field from 400 K to 5 K for samples S1-S3. The circle (1st), triangle (2nd) and square (10th) symbols hysteresis loops are shown in the figure 4. 4 (d)-(f) show the $H_{EB}$ plotted as a function of loop number $n$ (training effect).

In order to understand the relaxation behaviour of the interface spins, first we fitted the $H_{EB}$ vs $n$ data with the following empirical power law formula [30];

$$H_{EB}(n) = H_{EB\infty} + K/\sqrt{n} \qquad\qquad\qquad \text{................ (1)}$$

Where, $H_{EB}$ (n) is the exchange bias field for the nth loop run, $H_{EB\infty}$ is the exchange bias field after infinite number of loops run, $K$ is the system dependent constant. We found that the power law given in eq. (1) can fit well the experimental data for n>1 only and cannot fit the steep relaxation data between n=1 and 2. The $H_{EB\infty}$ and $K$ values show a systematic increase with decreasing the $Ir_{20}Mn_{80}$ layer thickness. Next, we considered a SG approach [11] containing both the frozen and rotatable interfacial spins of the FM/AFM system. We fitted the $H_{EB}$ vs $n$ data for all the samples by the double exponential decay function [11] which is specific for SG like frustrated interface:

$$H_{EB}(n) = H_{EB\infty} + A_f \exp(-n/P_f) + A_i \exp(-n/P_i) \qquad\qquad \text{............(2)}$$

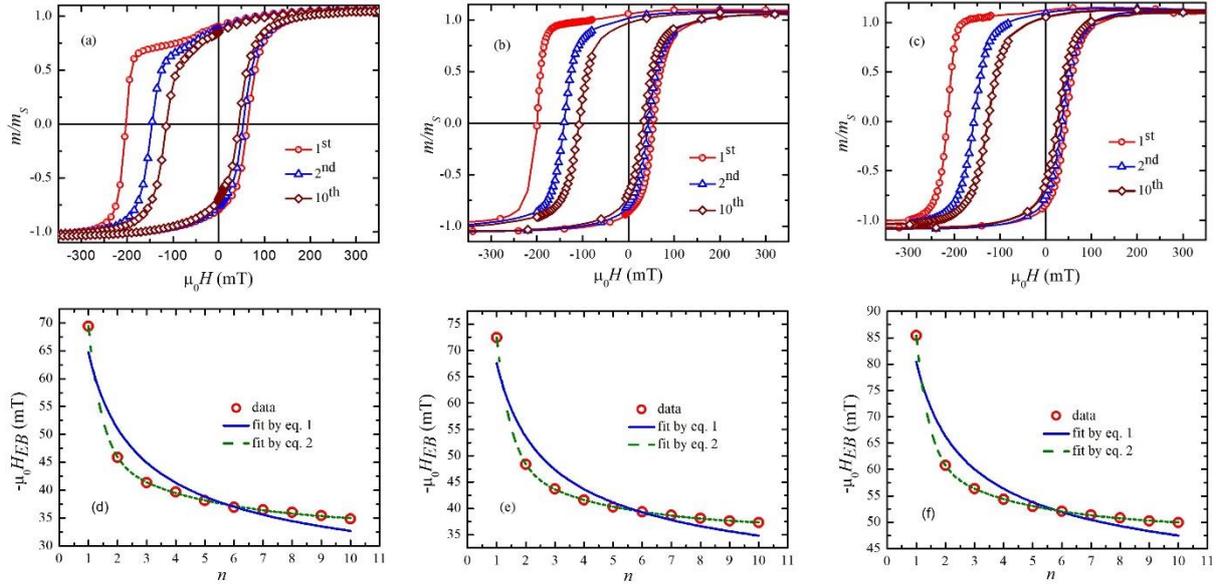

Figure 4: 1st, 2nd and 10th subsequent hysteresis loops of (a) S1, (b) S2 and (c) S3; variation of $H_{EB}$ vs $n$ for (d) S1, (e) S2 and (f) S3. In figures (d), (e) and (f), the open circles are experimental data, solid and segment lines are the fitted data.

Where, $A_f$, $P_f$ are the parameters for the interfacial frozen spins and $A_i$, $P_i$ are the parameters for the interfacial rotatable spins of the FM and AFM interface. $A_f$, $A_i$ have the dimension of mT and $P_f$, $P_i$ are dimensionless. $P_f$ and $P_i$ are the relaxation rates of the frozen and rotatable interfacial spin components, respectively. From the $R^2$ value, we can estimate the excellent fit of the training effect experimental data. The parameters obtained after fitting the training effect data with eq. (1) and (2) are given in table II. It is found the the fitting parameter $H_{EB\infty}$ shows a systematic decrease from 48.9 mT to 33.8 mT with increasing the thickness of

$Ir_{20}Mn_{80}$ from 3 to 10 nm. The weighing factors $A_i$ shows a systematic decrease with increasing $Ir_{20}Mn_{80}$ film thickness. However, the weighing factor $A_f$ value shows a little discrepancy in sample S2. The weighing factor $A_f$ is always higher than the $A_i$ indicating that the training effect is mainly due to the frozen spin component [6]. But we found that the relaxation rates of the frozen and rotatable interfacial spin components $P_f$ and $P_i$ remain constant for all the three samples. We calculated that the ratio $P_i/P_f$ for all the samples remains at 8. Thus, the relaxation rate of the interfacial rotatable spin component is almost 8 times faster than the interfacial frozen spin component. We can conclude that by considering both the interfacial frozen and rotatable spin components one can fit the experimental training effect data satisfactorily.

Table 2: fitting parameters obtained from the $H_{EB}$ vs $n$ plot for all the samples by fitting eq. (1) and (2).

| Sample | Equation (1) parameters | | Equation (2) parameters | | | | | |
|---|---|---|---|---|---|---|---|---|
| | $H_{EB-\infty}$ (mT) | $K$ (mT) | $H_{EB-\infty}$ (mT) | $A_f$ (mT) | $P_f$ | Ai (mT) | $P_i$ | $P_i/P_f$ |
| S1 | 17.9 ± 2.6 | 46.8 ± 4.8 | 33.8 ± 0.6 | 206.4 ± 34.3 | 0.46 ± 0.04 | 15.8 ± 1.6 | 3.93 ± 0.7 | 8.54 |
| S2 | 19.7 ± 2.7 | 47.9 ± 5.0 | 36.4 ± 0.2 | 199.5 ± 11.5 | 0.47 ± 0.02 | 16.0 ± 0.8 | 3.56 ± 0.2 | 7.57 |
| S3 | 32.3 ± 2.8 | 48.1 ± 5.2 | 48.9 ± 0.2 | 243.5 ± 16.6 | 0.43 ± 0.02 | 16.5 ± 0.7 | 3.58 ± 0.2 | 8.33 |

In the FM/AFM system the relaxation can be due to thermal or athermal effects [6]. The thermal training effects results into the gradual decrease of $H_{EB}$ with number of loops $n$. Due to the thermal activation, the AFM uncompensated spins are reconfigured from the original configuration [6]. In athermal training effect, there is a large reduction in $H_{EB}$ and $H_C$ between the first and second subsequent hysteresis loops afterwards a gradual decrease in $H_{EB}$ and $H_C$ is found. This type of relaxation arises due to the metastable state of the AFM spins [6]. We can conclude that the interfacial SG like structure frozen into a metastable stable state during FC and that metastable state relaxes resulting into the large reduction in the $H_{EB}$ and $H_C$ values in the second subsequent loop [6].

Thus, this study develops our general understanding in the role of interfacial SG state on the magnetic properties. Again, the training effect data tells that both the frozen and rotatable spins have effect in EB. Also, we investigated the relaxation rate of rotatable spins wrt the frozen spins.

In conclusion, we investigated the temperature, $H_{FC}$ dependence of EB and training effect in sputter-deposited $Fe/Ir_{20}Mn_{80}$ systems. We observed exponential decay behaviour in $H_{EB}$, $H_C$ with temperature. We found decrease in $H_{EB}$ with increase in $H_{FC}$ contrary to the typical behaviour $H_{EB}$ with $H_{FC}$ in FM/AFM systems. It is found that the SG like interface might be a primary reason for this type of trend in $H_{FC}$ and temperature dependence of EB. However, the contribution of 'bulk' part of the AFM spins to EB from the variation of $H_{EB}$ with the thickness of AFM is also observed. Training effect has been analysed by fitting a model which consider both frozen and rotatable spins. We found that the interfacial rotatable spins relax 8 times faster than the interfacial frozen spins.

## Supporting Information

Supporting information file 1: the GIXRD (Figure S1) data of all the samples.
File Name: S1.pdf
Format: pdf

## Acknowledgements

The authors thank Department of Atomic Energy (DAE) and Department of Science and Technology-Science and Engineering Research Board (SB/S2/CMP-107/2013), Government of India for providing the financial support.